\documentclass[superscriptaddress, twocolumn, prl]{revtex4}
\usepackage{graphicx}   
\usepackage{array}
\usepackage{array}
\usepackage{bibunits}
\newcolumntype{x}[1]{>{\centering\arraybackslash\hspace{0pt}}p{#1}}

\newcommand{\figref}[1]{\figurename~\ref{#1}}

\newcommand{\meter}[1][]{\ifx|#1|\unit{m}\else\unit[#1]{m}\fi}
\newcommand{\hertz}[1][]{\ifx|#1|\unit{Hz}\else\unit[#1]{Hz}\fi}
\newcommand{\fm}[1][]{\ifx|#1|\unit{fm}\else\unit[#1]{fm}\fi}
\newcommand{\fluence}[1][]{\ifx|#1|$\unit{mJ/cm^2}$\else$\unit[#1]{mJ/cm^2}$\fi}

\defaultbibliography{M:/PostDoc_FEMTO/Jabref_files/references_Heusler}
\defaultbibliographystyle{apsrev}

\begin{document}
\begin{bibunit}
\title{Structural and Magnetic Dynamics in the Magnetic Shape Memory Alloy Ni$_2$MnGa}

\author{S. O. Mariager}
\email{simonforscience@gmail.com}
\affiliation{Swiss Light Source, Paul Scherrer Institut, 5232 Villigen, Switzerland}
\author{C. Dornes}
\affiliation{Institute for Quantum Electronics, ETH Zurich, 8093 Z\"urich, Switzerland}
\author{J. A. Johnson}
\affiliation{Swiss Light Source, Paul Scherrer Institut, 5232 Villigen, Switzerland}
\author{A. Ferrer}
\affiliation{Institute for Quantum Electronics, ETH Zurich, 8093 Z\"urich, Switzerland}
\author{S. Gr\"{u}bel}
\affiliation{Swiss Light Source, Paul Scherrer Institut, 5232 Villigen, Switzerland}
\author{T. Huber}
\affiliation{Institute for Quantum Electronics, ETH Zurich, 8093 Z\"urich, Switzerland}
\author{A. Caviezel}
\affiliation{Swiss Light Source, Paul Scherrer Institut, 5232 Villigen, Switzerland}
\author{S. L. Johnson}
\affiliation{Institute for Quantum Electronics, ETH Zurich, 8093 Z\"urich, Switzerland}
\author{T. Eichhorn}
\affiliation{Institut f\"{u}r Physik, Johannes Gutenberg Univ. Mainz, 5128 Mainz, Germany}
\author{G. Jakob}
\affiliation{Institut f\"{u}r Physik, Johannes Gutenberg Univ. Mainz, 5128 Mainz, Germany}
\author{H. J. Elmers}
\affiliation{Institut f\"{u}r Physik, Johannes Gutenberg Univ. Mainz, 5128 Mainz, Germany}
\author{P. Beaud}
\affiliation{Swiss Light Source, Paul Scherrer Institut, 5232 Villigen, Switzerland}
\author{C. Quitmann}
\affiliation{Swiss Light Source, Paul Scherrer Institut, 5232 Villigen, Switzerland}
\author{G. Ingold}
\email{gerhard.ingold@psi.ch}
\affiliation{Swiss Light Source, Paul Scherrer Institut, 5232 Villigen, Switzerland}


\begin{abstract}
Magnetic shape memory Heusler alloys are multiferroics stabilized by the correlations between electronic, magnetic and structural order. To study these correlations we use time resolved x-ray diffraction and magneto-optical Kerr effect experiments to measure the laser induced dynamics in a Heusler alloy Ni$_2$MnGa film and reveal a set of timescales intrinsic to the system. We observe a coherent phonon which we identify as the amplitudon of the modulated structure and an ultrafast phase transition leading to a quenching of the incommensurate modulation within 300~fs with a recovery time of a few ps. The thermally driven martensitic transition to the high temperature cubic phase proceeds via nucleation within a few ps and domain growth limited by the speed of sound. The demagnetization time is 320~fs, which is comparable to the quenching of the structural modulation.
\end{abstract}

\maketitle

Multiferroic materials which exhibit large responses to electromagnetic and stress fields are of great interest for novel technical applications. To guide their rational design the microscopic origin of their functional properties must be understood which requires methods that can disentangle the interplay between electronic, magnetic and structural degrees of freedom. A prototypical example is the optimization of ferromagnetic X$_2$YZ Heusler alloys, where one class shows novel functional properties such as magnetic shape memory and magnetocaloric effects due to the coexistence of ferromagnetism and a structural martensitic (MT) transition \cite{Planes2009}, and another class is half-metallic and suitable for spintronic applications \cite{Graf2011}. Ni$_2$MnGa is the classical Heusler magnetic shape memory alloy with a magnetically induced strain of up to 10\% arising from the interplay between magnetic and structural domains in the twinned low temperature MT phase \cite{Ullakko1996,Murray2000,Sozinov2002,Planes2009}. The structure of the MT phase changes with alloy composition \cite{Lanska2004}, but the modulated phases (commonly labelled 5M and 7M) \cite{Righi2007,Righi2008} displaying magnetic shape memory only exist if the MT transition temperature $T_{MT}$ is lower than the Curie temperature $T_C$ \cite{Lanska2004,Mariager2014}. In these structures the minimum in free energy is shifted such that the lattice constant ratio is $c/a<1$ in the splitting (tetragonal or orthorhombic) of the high temperature cubic austenite (AUS) phase, compared to the usual global minimum found at $c/a > 1$ in the non-modulated tetragonal phase. Theoretical studies suggest that the modulation of the structure stabilizes this new minimum \cite{Bungaro2003,Zayak2004,Entel2013}. In Ni$_2$MnGa compounds, we then have a situation where the interplay between an incommensurate structural modulation, a splitting of the electronic states due to the tetragonal or orthorhombic distortion of the cubic lattice and the ferromagnetic order combine to stabilize a phase displaying a large magnetic shape memory effect.

In this letter we study this interplay by employing ultrafast time resolved x-ray and optical methods to separate the three types of order in time and to investigate the possible coupling between ferromagnetism and the modulated structure. We measure a satellite reflection sensitive to the modulated structure, a (202) Bragg reflection sensitive to the orthorhombic splitting, and optically, the change in the magnetization. This allows us to separate the fast electronic processes from thermally driven ones as we, upon photoexcitation, follow the evolution from the low temperature MT phase to the high temperature AUS phase. We quantify the relevant phonon modes and intrinsic time scales and show that while the modulation of the structure and the magnetization are quenched on the same timescale, they have different recovery times.

The 1~$\mu$m thick Ni$_{2.085}$Mn$_{1.133}$Ga$_{0.782}$ film was grown by dc-magnetron sputtering on a MgO(100) substrate with a 100~nm Chrom buffer layer \cite{Jakob2007, Eichhorn2011}. The Curie temperature is $T_C$ = 368~K and the transition temperature $T_{MT}$ measured as the average of onset and completion is 349~K upon cooling and 354~K upon heating. The crystal structure of the MT phase was measured at the Material Science beamline of the Swiss Light Source \cite{Willmott2013} to be orthorhombic with MT twinning leading to 12 different MT domains as observed in similar films \cite{Khachaturyan1983, Klaer2011,Eichhorn2012}. The MT structure is modulated \cite{Zheludev1996,Righi2007,Mariager2014} but unlike the single modulation wavevector $\mathbf{q} = [\xi\xi0]$ with $\xi = 0.428$ of single crystal Ni$_2$MnGa, the film has two modulation wavevectors with $\xi = 0.311(2)$ and $0.51(1)$ in reciprocal lattice units of the orthorhombic lattice. The first wavevector is close to the $\xi=0.308$ found for the 7M structure \cite{Righi2008}.

The time-resolved x-ray diffraction was performed at an x-ray energy of 5~keV using a synchrotron slicing source (200~ph/pulse, 2~kHz and 1.2 \%bw) as probe and an 800 nm, 120 fs 1~kHz \textit{p}-polarized laser pulse at $12^o$ incidence as pump (absorption length $\delta = 20$~nm \cite{Zhou2002}), giving a total time resolution of 200~fs \cite{Beaud2007}. The x-ray spot was focused horizontally to 400$~\mu$m, vertically to 10$~\mu$m and the gracing incidence angle was set to $0.65^o$ (absorption length $\alpha = 20$~nm \cite{Henke1993}) to match the penetration depth of the x-ray probe to the laser pump. The experiments were carried out at a sample temperature of 340~K.

The time resolved magneto-optical Kerr (tr-MOKE) effect was measured in a polar geometry with the pump laser at normal incidence ($90^{\circ}$), the probe at $60^{\circ}$ and a 0.6~T magnetic field. The 80~fs 800~nm pump and probe beams were cross polarized to minimize pump scatter. The detection was done with two balanced photodiodes and the signal was enhanced by measuring at a small heterodyne angle. The MOKE signal is found as the difference between two opposite magnetic fields and is proportional to the out of plane magnetization $m_z$ if in-plane dynamics and second order terms are ignored \cite{You1996}.

\begin{figure} [ht]
\includegraphics[scale=.55]{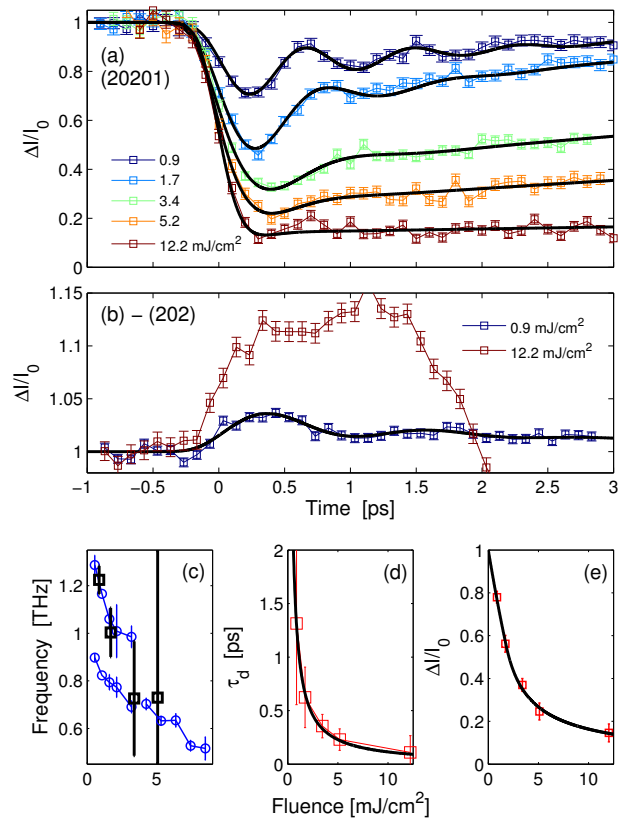}
\caption{(Color online) Modulation dynamics. (a) Change in intensity of the (20201) satellite reflection as a function of time after laser excitation at five laser pump fluences. (b) Change in intensity of the MT (202) Bragg reflection. The color code matches (a). (c) Oscillation frequencies from optical data (circles) and the x-ray data (black squares) from (a) as a function of laser fluence. (d) Damping times from the fits in (a). (e) Intensity of the (20201) reflection after $\sim500$~fs (red squares). The black line is a fit as described in the text.} \label{figAmplitudon}
\end{figure}

In \figref{figAmplitudon}(a) we present the dynamics of the (20201) reflection, where the two extra indices label the first satellite of the (202) reflection, arising from the second modulation. At low laser pump fluences (0.9~mJ/cm$^2$ absorbed) we observe a 1.2~THz oscillation caused by a coherent optical phonon. As the fluence is increased both the amplitude and damping of the oscillation increase, and a recovery of the intensity on a picosecond timescale is evident. At high fluence (12.2~mJ/cm$^2$) a fast drop of the intensity occurs within $\tau_e=300\pm40$~fs which roughly corresponds to a quarter of the fastest measured phonon period. At these high fluences the disappearance of the peak is a clear sign that the structural modulation is gone and that an ultrafast phase transition resulting in an increase in crystal symmetry has occurred. The fact that the intensity does not drop completely to zero can be attributed to the finite x-ray and laser penetration depths.

A structural change solely due to a change in the modulation amplitude identifies the observed coherent phonon as the amplitudon, for which the corresponding phason mode has previously been observed by neutron scattering \cite{Shapiro2007}. This interpretation is supported by the measurement of the MT (202) Bragg reflection shown in \figref{figAmplitudon}(b). On short timescales the intensity of the (202) reflection increases, indicating an increase in orthorhombic order consistent with the suppression of the modulation. At low fluence (0.9~mJ/cm$^2$) a very weak oscillation appears. The drop in intensity after 1.5~ps for the data taken at 12.2~mJ/cm$^2$ is due to the thermal expansion and the MT to AUS transition.

To quantify the observed oscillation the data of the (20201) peak was fitted with the function $I(t)/I_0 = |1+A(\cos(2\pi\nu t)e^{-t/\tau_d}-1)e^{-t/\tau_r}|^2$, convolved with a gaussian function accounting for the time resolution. This function with cosine phase describes a displacive excitation \cite{Zeiger1992} resulting in a damped ($\tau_d$) oscillation around a new equilibrium position (1-A). The second exponential accounts for the observed recovery ($\tau_r$). This choice for I(t) assumes that the x-ray structure factor is proportional to the modulation amplitude and accounts for the intensity being proportional to the square of the structure factor.

In \figref{figAmplitudon}(c) the frequencies obtained from the fits are plotted together with the frequencies obtained from time resolved optical reflectivity measurements. At low fluence the x-ray and optical data are in good agreement for one of the two observed optical phonon branches while frequencies cannot reliable be extracted from the x-ray data at higher fluences. In stoichiometric Ni$_2$MnGa single crystals we observed only a single frequency in optical experiments \cite{Mariager2012a}, in agreement with the existence of a single modulation wavevector \cite{Mariager2014}. In the present film sample we observe two modulation wavevectors and two frequencies. While the frequencies correspond to the excitation of two amplitudons, as observed the (20201) reflection is only sensitive to one of these. On the other hand, the (202) peak should be sensitive to both the 0.8 and 1.2~THz oscillations, again consistent with the data.

To support the model of an amplitudon we plot the change in intensity of the (20201) reflection immediately after excitation in \figref{figAmplitudon}(e), extracted from the fits as $\Delta I/I_0 = (1-A)^2$. The initial linear drop in intensity vs fluence ($<2$~mJ/cm$^2$) indicates a linear decrease in modulation amplitude with laser fluence, while the gradual drop at higher fluences is due to the comparable x-ray and laser penetration depths. To illustrate this we fit the fluence curve with a simple model. We assume that the transient equilibrium modulation amplitude x is linear in laser fluence for $f<f_c$ and $x$ = 0 for $f\geq f_c$. As a function of depth $z$ in the sample we write $x(z) = x_0(1-f(z)/f_c) = x_0(1-f_0e^{-z/\delta}/f_c)$ for $f(z)<f_c$ where $x_0$ is the amplitude before excitation. We then calculate the lattice sum over the lattice planes into the crystal as $S = \sum_{n_z}F(x(z))e^{2\pi i ln_z}e^{-n_za/2\alpha}$, with the measured intensity given as $I = |S|^2$. Here $l$ = 2 is the out of plane reciprocal lattice coordinate and F is the unit cell structure factor which for a satellite reflection of an incommensurate modulation is proportional to a first order Bessel function, $F(x) = (-1)^1J_1(x)$ \cite{Smaalen2007}. The fit then only depends on the ratio of the x-ray to laser penetration depths $\alpha/\delta$ and the critical incident fluence $f_c$. As seen in \figref{figAmplitudon}(e) the fit describes the data very well ($R^2 = 0.997$) with $\alpha/\delta = 1.41\pm0.01$ and $f_c = 1.83\pm0.01$~mJ/cm$^2$. We note that the fit is independent of the initial value $x_0$ in the linear regime of the Bessel function ($x_0 < 0.2$). For comparison several other models for x(z) have been tested as outlined in the supplementary material \cite{Supplementary}. To conclude the model shows how the transient equilibrium amplitude of the modulation changes linearly with the laser excitation, which leads to a displacive excitation of the amplitude mode.

The oscillations are absent at high fluences due to the strong damping, which is shown as a function of fluence in \figref{figAmplitudon}(d). For the observed fluence range the damping time $\tau_d$ is inversely proportional to the fluence, as illustrated by the fit ($\tau_d = d/f$, $d = 1.14\pm0.05$~ps).

\begin{figure}[t]
\includegraphics[scale=.55]{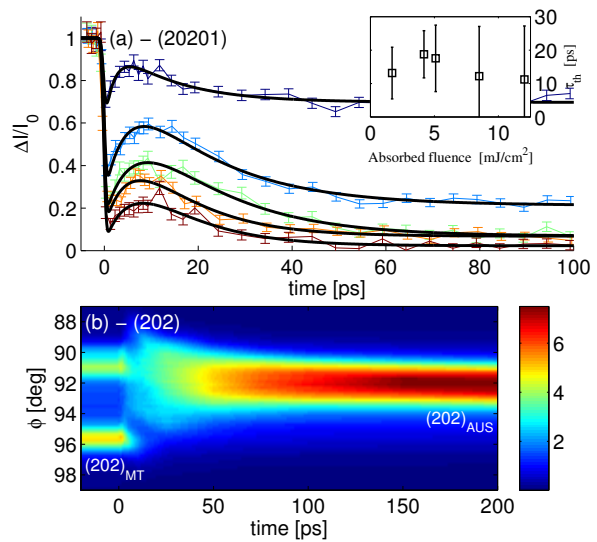}
\caption{MT to AUS transition. (a) Change in intensity of the (20201) satellite reflection as a function of time and laser fluence. The insert shows the thermal transition time $\tau_{th}$ as a function of laser fluence. (b) Color plot of the dynamics of the (202) lattice reflection.} \label{figSlLong}
\end{figure}

The ultrafast electronic quenching of the modulation is distinctly different from the slower thermally driven martensite transition. For comparison we show in \figref{figSlLong}(a) the evolution of intensity of the (20201) reflection up to 100~ps. On this timescale the initial coherent dynamics are only visible as an immediate drop in intensity followed by a gradual recovery. This recovery is consistent with a thermalization of the electrons and the lattice leading to a restoration of the potential which give rise to the modulation. On longer timescales a second drop in intensity occurs, which at high fluences leads to a complete quenching of the modulation. To extract the timescales the data was fitted using $I(t) = 1-A_1e^{-t/\tau_r}-A_2(1-e^{-t/\tau_{th}})$ accounting for the initial drop ($A_1$), the recovery ($\tau_r$) and a second exponential drop ($A_2,\tau_{th}$). The weighted mean of the recovery times is $\tau_r = 4\pm1.7$~ps. The transition time $\tau_{th}$ shown in the inset in \figref{figSlLong}(a) does not depend on the excitation fluence and has a mean value of $\tau_{th} = 15\pm3$~ps. Due to the short x-ray absorption length the exponential drop in intensity is expected for a transition moving into the crystal at a constant speed, and nucleated MT domains are known to grow at the speed of sound \cite{Khachaturyan1983}. The speed of sound in the [100] direction of the modulated structure is estimated to be $v \approx 5$~nm/ps \cite{Li2011}. A phase transition moving into the crystal at this speed would produce an exponential change in intensity with a time constant of $\sim7$~ps. Accounting for the thermalization time between the electrons and the lattice we conclude that nucleation occurs mainly within the first 10 ps after excitation. The existence of the intermediate recovery however shows that the initial electronic quenching does not directly launch the final thermal phase transition.

The change in intensity of the satellite reflection can be compared to the change of the regular lattice peaks shown in \figref{figSlLong}(b). Rocking curve scans obtained at different time delays show how the (202) MT peaks transform to a single AUS (202) reflection at a pump fluence of 12.2~mJ/cm$^2$. The fast disappearance of the MT peaks is best seen at $\phi = 95.7^{\circ}$, where we extract a timeconstant of $\tau_{MT} = 14\pm4$~ps, in agrement with the time $\tau_{th}$ found for the (20201) reflection. That is, on this thermal timescale the disappearance of the MT splitting and the incommensurate modulation proceed together. The rising AUS lattice peak ($\phi = 92^{\circ}$) exhibits significantly slower dynamics with a rise time of 200~ps. This is due to the fact that the thin top layer directly probed by the x-rays is oriented and strained by the lower part of the film. The transformation of the entire film is limited by the speed of sound, leading to a timescale of approximately $d/v \approx$ (1~$\mu$m)/(5~nm/ps) = 200~ps.

\begin{figure} [t]
\includegraphics[scale=.55]{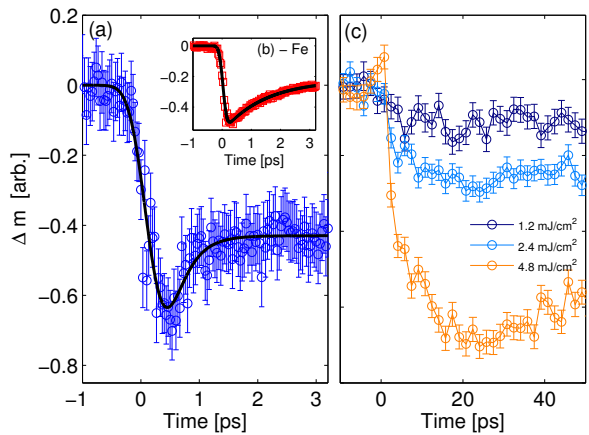}
\caption{Magnetization dynamics. (a) Change in magnetization as a function of time for Ni$_2$MnGa. The pump fluence was 1.4~mJ/cm$^2$. (b) Reference trace taken on an Fe(001) single crystal. (c) Change in magnetization for Ni$_2$MnGa at times up to 50~ps for three different fluences.} \label{figMoke}
\end{figure}

As pointed out in the introduction, the modulated structures only exist if $T_{MT} < T_C$. This raises the question if the modulation is stabilized by an interaction between structural and magnetic order. To explore this we measured the demagnetization time after laser excitation. The tr-MOKE data at times up to 3~ps are shown in \figref{figMoke}(a), with a reference trace measured on an Fe(001) single crystal shown for comparison in the insert. In both traces we observe a fast drop usually assigned to the demagnetization \cite{Kirilyuk2010,Malinowski2008}, as well as a partial recovery of the signal. The Fe trace shows a demagnetization time of $140\pm20~fs$ and is in good agreement with previous experiments \cite{Kampfrath2002}. In Ni$_2$MnGa we find a demagnetizaiton time of $\tau_m = 320\pm50~fs$. Both demagnetization times were extracted by fitting the data with a step function combined with an exponential recovery. The Ni$_2$MnGa demagnetization time is in agreement with those found for Heusler alloys with similar spin polarization factors, though these were measured in spintronic compounds with Aus structures \cite{Steil2010,Mann2012}. We thus find that the demagnetization and the quenching of the structural modulation occur essentially simultaneously. If we instead measure the time evolution up to 50~ps as shown in \figref{figMoke}(c) the tr-MOKE data does not show an intermediate recovery, unlike the structural data in \figref{figSlLong}(a). We conclude that the magnetization and the structural modulation have different dynamics after the initial drop. The modulation therefor then can not be stabilized by a direct coupling to the magnetic moment, though they are both rapidly quenched after the laser excitation of the electronic subsystem.

Upon laser excitation several timescales are present in Ni$_2$MnGa. The structural modulation is fully quenched within 300~fs, about a quarter of the corresponding phonon period. The ferromagnetic order is quenched with a demagnetization time of 320~fs and the MT to AUS transition occurs after thermalization of the electron and lattice ($\sim 4$~ps) with nucleation within 10~ps and the subsequent domain growth limited by the speed of sound. These timescales are independent of the excitation fluence and intrinsic to the system. The fact that the ultrafast structural transition appears before thermalization between the electrons and the lattice, and recovers upon thermalization despite the demagnetization indicates that the origin of the incommensurate modulation is the strong electron phonon interaction. A possible explanation is the destabilization of the incommensurate modulation by distortion of the Kohn anomaly \cite{Bungaro2003} near the Fermi edge via excitation of Ni $3d$ electrons to unoccupied states just above the Fermi edge. Such states are available in both minority and majority bands \cite{Barman2005,Opeil2008} and the importance of Ni $3d$ states for the structural transitions in Ni$_2$MnGa has previously been outlined \cite{Jakob2007}. An alternative mechanism could involve destruction of the suggested hybrid bonding between Ni d and Mn p states near the Fermi level \cite{Zayak2004,Kubler1983}. On the other hand the quenching of both the modulation and the magnetism does not lead directly to the MT transition. The transition only occurs when a temperature above $T_{MT}$ has been reached and all subsystems are in thermal equilibrium.

To summarize, we measured a coherent phonon mode and an ultrafast non-thermal structural transition in an incommensurately modulated Ni$_2$MnGa film. Both the observation of this phase transition and the displacive excitation of the phonon is consistent with the concept of an amplitudon. The dynamics of this electronically driven ultrafast transition is distinctly different from the much slower thermally driven martensite transition, which proceeds with nucleation within 10~ps and domain growth limited by the speed of sound. The different recoveries of the demagnetization and the structural modulation indicates that the origin of the modulation is electronic. This examples shows how ultrafast pump-probe methods applying a variety of specific probes can be used to disentangle the different degrees of freedom in multiferroic materials.

\begin{acknowledgments}
The x-ray experiments were performed on the X04SA and X05LA beamlines of the Swiss Light Source, Paul Scherrer Institut, Villigen, Switzerland. We thank P. Willmott, D. Grolimund and C. Borca for help. This work was supported by the Swiss National Foundation through NCCR MUST, by the DFG through SPP 1239 within Ja821/3-3 and by the Graduate School of Excellence “Materials Science in Mainz” (MAINZ).
\end{acknowledgments}

\putbib
\end{bibunit}

\begin{bibunit}

\newpage
\appendix
\onecolumngrid
\begin{centering}
\Large
\textbf{Supplementary Material} \\
\end{centering}

\section{Structure}
The static x-ray diffraction experiments were performed at the Material Science beamline of the Swiss Light Source, with a (2D+2S) surface diffractometer equipped with a PILATUS 100K two-dimensional detector \cite{Willmott2013}. The 2D detector allowed us to sample large three-dimensional reciprocal space volume data sets which were converted to I(h,k,l), intensity as a function of coordinates in reciprocal space \cite{Schleputz2011}. From these data reciprocal space maps could subsequently be extracted. The x-ray energy was 17~keV and the x-ray incidence angle was kept constant to ensure a homogeneous footprint. The sample temperature was controlled using a nitrogen cryojet, and the transition temperatures were in agreement with magnetization measurements. The sample was initially oriented in the AUS phase, and the resulting orientation matrix was used as reference when cooling to the MT phase.

\begin{figure}[ht]
\includegraphics[scale=.7]{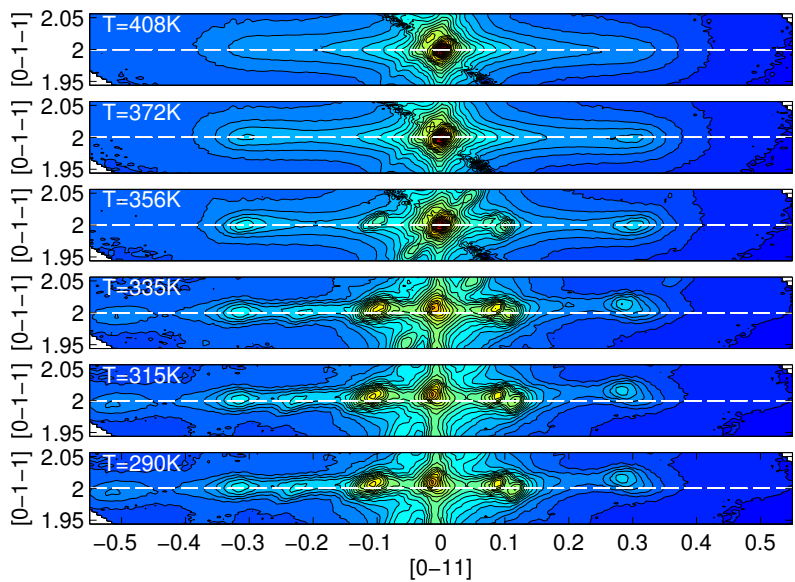}
\caption{(Color online) A series of reciprocal space maps taken at different temperatures during cooling. The premartensite satellite reflection at $\pm.42$ (which is .33 in r.l.u.) slowly appears from T = 408~K where it is hardly visible towards 356~K. At lower temperature the martensite transition has occurred and new satellite lattice peaks appear. The intensity of these peaks are independent of temperature (T = 315 and 290~K) and are clearly not located exactly on the [01$\overline{1}$]$_A$ rod.} \label{figrsm_T2}
\end{figure}

In \figref{figrsm_T2} we show reciprocal space maps through the $(202)_{AUS}$ reflection as the sample is cooled from  408~K to 290~K. As the temperature is lowered we first (T=372~K and 356~K) see the appearance of the satellite reflections from the pre-martensite phase \cite{Zheludev1996,Mariager2014}, which has a modulation wavevector of $\mathbf{q} = [\xi\xi0]$ with $\xi \approx 1/3$. As the MT transition is reached (T = 356~K and 335~K) the regular lattice peak is split and several Bragg reflections corresponding to different MT twins can be seen, though none coincide exactly with the shown lattice plane.

\begin{figure*}[hbt]
\includegraphics[scale=.68]{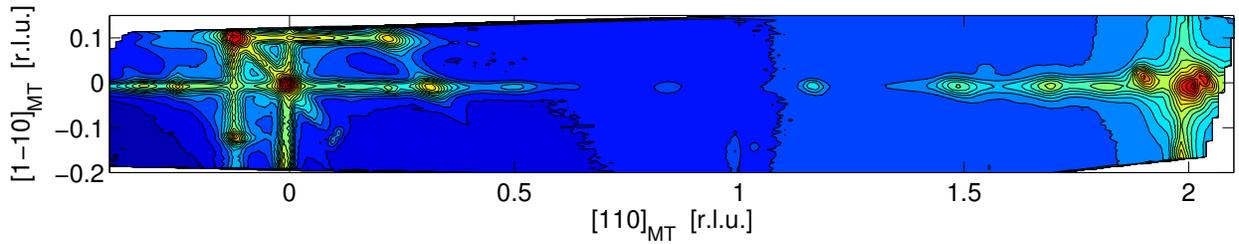}
\caption{(Color online) Reciprocal space map showing the streaks of scattering
extending from a [202]$_{MT}$ to a [224]$_{MT}$ Bragg peak. It shows the $c_{mt}$ plane and in AUS coordinates corresponds to a cut through [2.103 0.105 2.015] with surface normal [1.0568 0.0547 0].} \label{figctr2}
\end{figure*}

To identify the satellite reflections from the structural modulation of the MT phase, a cut has to be made between two MT peaks corresponding to the same twin. One example of such a reciprocal space map can be seen in \figref{figctr2}. It shows a cut including the $[202]_{MT}$ and $[224]_{MT}$ peaks corresponding to a single structural twin and can be used to identify the many satellite reflections located exactly between them.

\begin{figure}[ht]
\includegraphics[scale=.7]{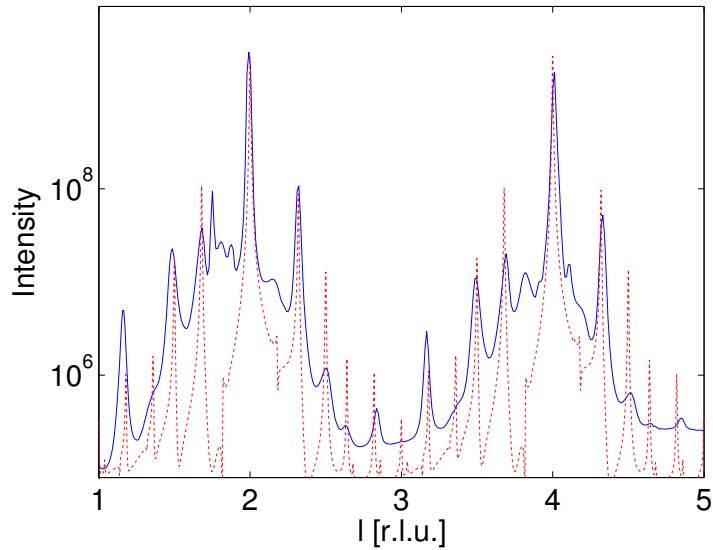}
\caption{(Color online) A line scan taken between two [0-22] and [2-24] peaks of a single MT twin. The rich structure in the immediate vicinity of the regular reflections (l = 2 and l = 4) arise from the other 11 twins. The red dotted line shows the calculated peak positions from a structure with two modulation periods.} \label{figmod1}
\end{figure}

In \figref{figmod1} we show an example of a line scan along such a rod. We note that this scan is representative, that is, if there are any satellite reflections between two Bragg reflections corresponding to a single twin, we observe exactly these peaks. This structure is somewhat similar to scans published by \textcite{Eichhorn2011}, where they attributed it to a 7M structure. We observe the following concerning the location of the satellite reflections. First, all peaks appear modulo 2 r.l.u. and in pairs at $\pm q$ from the main reflections. In addition they are not placed at periodic intervals between the reflections with the main reflections occurring at $\Delta q = \pm 0.311(1), 0.51(1), 0.83(1)$. This is significantly different from the incommensurately modulated structure known from stochiometric Ni$_2$MnGa \cite{Righi2007} and it is not consistent with the typical 5M or 7M phases in any of their manifestations. The symmetric reflections indicate a periodic modulation along the \{110\} axes of the sample, but the different deviations $\Delta q$ indicate that more than a single modulation period exists. In fact, the observed reflections are consistent with two wavevectors with $\xi = 0.31$ and 0.51. Two such modulations will result in peaks at any combination of the modulation vectors, explaining the observed reflection at 0.83 (within the measurement precision). To test this we calculate the scattering from a structure with two such modulation periods \cite{Ustinov2009,Mariager2014} and as shown in \figref{figmod1} the result is a very good agreement between observed and calculated peak positions. The intensities are not reproduced as the calculation did not include the unit cell formfactor, allowed for different modulation amplitude for different atoms or included experimental factors such as polarization and the Lorentz factor.

We finally emphasize that the observed Bragg reflections are not compatible with an adaptive phase \cite{Kaufmann2010}. The satellite lattice reflections from such a structure would result in evenly spaced intermediate reflections, with only small possible deviations due to stacking faults \cite{Ustinov2009,Mariager2014}

\section{Optical data}
The photoinduced change in reflectivity was investigated in an optical pump-probe setup with a time resolution of $\sim 80$~fs. The 800~nm pump and probe pulse were generated with a 2~kHz Ti-Saphire laser system. The pump and probe pulses were cross polarized and incident along the [001] surface normal with the probe polarization aligned to the [100] crystal axis. The pump was focused to $500\times500~\mu m^2$ and the probe spot size was a factor of three smaller to ensure homogenous excitation of the probe region. The sample was mounted in a cryostat.

\begin{figure} [ht]
\includegraphics[scale=.8]{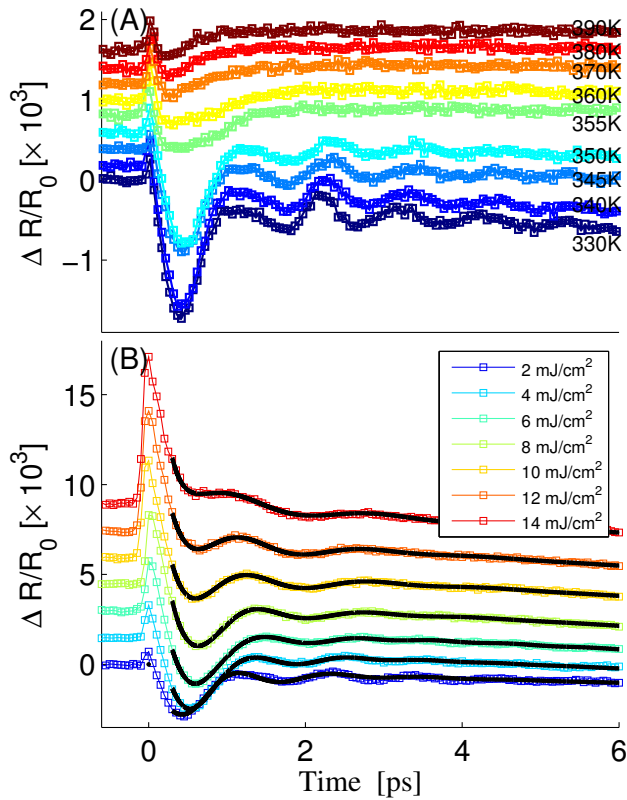}
\caption{All optical pump-probe experiments. (A) Time resolved reflectivity as a function of temperature at a fluence of 1~mJ/cm$^2$. (B) Time resolved reflectivity as a function of pump fluence. The black lines shows the fits used to extract the frequencies presented in the main text.} \label{figOpt}
\end{figure}

In \figref{figOpt}A we show the result of a time resolved optical pump-probe reflectivity measurements taken above and below $T_{MT}$ with a pump fluence of 1~mJ/cm$^2$. Below $T_{MT}$ the signal shows clear oscillations, with a beating indicating the presence of more than a single frequency. Above $T_{MT}$ no oscillations are present, beyond a single initial dip in reflectivity. In addition there is an initial electronic peak, whose amplitude as seen in \figref{figOpt}B scales with fluence.  The phase transition is clearly visible as the active phonon modes change with the change in lattice symmetry. This behavior is qualitatively consistent with what we have observed in single crystal Ni$_2$MnGa \cite{Mariager2012a}. Both a Fourier analysis and a fit including two cosine functions reveal the existence of two distinct frequencies, around 1.3 and 0.9~THz just below T$_{MT}$ and hardening as the temperature is lowered. Given the two modulation wavevectors of the structure, the two coherent phonons can be attributed to two amplitudons, that is, an oscillation of the modulation amplitude. This is consistent with the existence of an amplitudon in stochiometric single crystal Ni$_2$MnGa \cite{Mariager2012a}. In \cite{Mariager2012a} we could not distinguish between an amplitudon and a zone folded acoustic mode, but with the structure now absolutely determined as incommensurately modulated \cite{Mariager2014} the observed phonon can be assigned to the amplitudon.

\section{Amplitudon Model}
In the fit in FIG.1(e) of the main paper we assumed that the modulation amplitude varied linearly with the laser fluence. Several other models have been tested. An assumption that the modulation amplitude falls of as a power of f results in a significantly worse fits ($R^2 = 0.931$ for $x = x_0(1-\sqrt{f/f_c}$ )). On the other hand a model with a true threshold where $x = x_0$ for $f<f_c$ and 0 for $f\geq f_c$ cannot be ruled out ($R^2 = 0.990$) due to the low number of data points at low fluence and gives $f_c = 0.61(5)$~mJ/cm$^2$. This model is however complicated as it must be combined with the existence of a coherent phonon to account for the data in FIG.1(a) of the main text.

The critical fluence $f_c = 1.83(1)$~mJ/cm$^2$ can be converted to a corresponding temperature, though this number depends on the laser absorption depth. For $\delta = 20$~nm we find a critical energy density of 797~J/cm$^3$, which is about 1~eV per unit cell or a temperature rise of 250~K. While it then appears that the energy needed to quench the modulation on an ultrafast timescale is significantly higher than the thermal energy needed to reach $T_{MT}$, one explanation might be that this energy is distributed in various electronic channels from which only a fraction contributes to quenching the modulation.

\section{Time resolved rocking curves}
\begin{figure*} [ht]
\includegraphics[scale=.65]{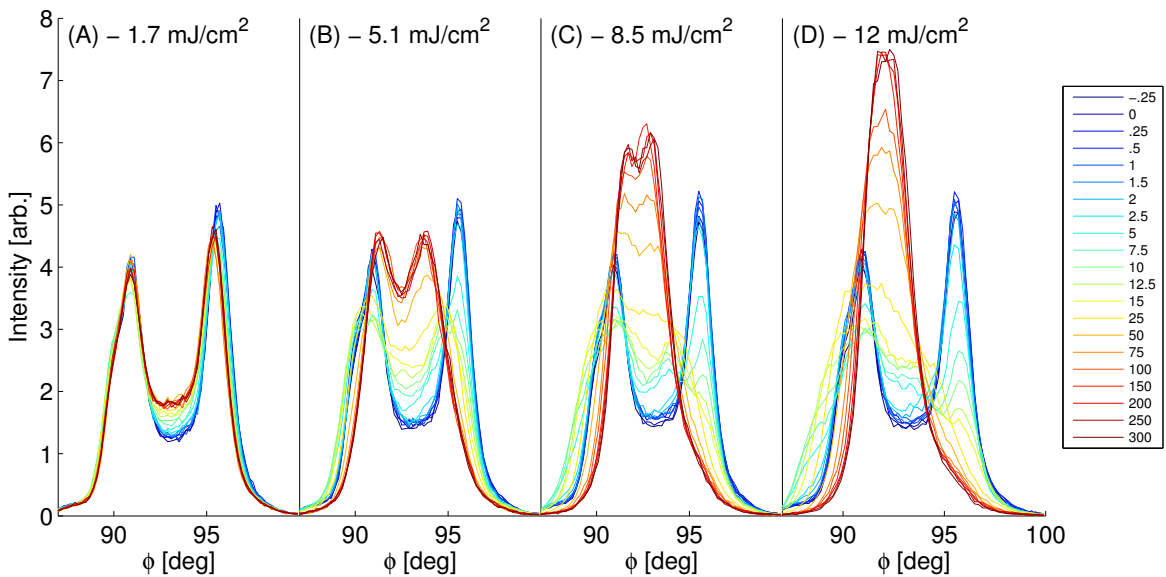}
\caption{(Color online) Rocking curves as a function of time delay at four different pump fluences. The legend shows the delay in units of ps, and is the same for all figures. (A) 1.7~mJ/cm$^2$, (B) 5.1~mJ/cm$^2$, (C) 8.5~mJ/cm$^2$, (D) 12~mJ/cm$^2$.} \label{figAusRC}
\end{figure*}

The data in FIG.2(b) of the main paper is composed from rocking curves (rotation of the sample around the surface normal) taken at different time delays. The rocking curves for four different pump fluences are shown in \figref{figAusRC}. The un-pumped rocking curve, as seen in \figref{figAusRC}A at t = -0.25~ps is, a combination of the rocking curves from 6 of the 12 MT twins, which are here probed in a single scan. The shape can be reproduced from the static x-ray data if one take into account the opening angles of the avalanche photo diode used for the time-resolved experiments. At a low fluence of 1.7~mJ/cm$^2$ the transition is only seen as a small increase in intensity at the location of the AUS peak, while the majority of the film remains in the MT phase. Only at the high fluence of 12~mJ/cm$^2$ in \figref{figAusRC}D is the high temperature AUS phase reached. Here the entire film is transformed and a single strong AUS (202) Bragg reflection occurs at $\Phi = 92.1^{\circ}$. At the intermediate fluences only parts of the film is transformed, resulting in a strained AUS structure.

The temperature rise due to a single laser pulse is given as:
\begin{equation}
\Delta T = (1-R)f_0(1-e^{-d/\delta})/C\rho d
\end{equation}
The specific heat of Ni$_2$MnGa is C = 0.35 - 0.4~J/gK and the density is $\rho = 8.14~g/cm^3$. The laser penetration depth at $\lambda=800$~nm is $\delta=20~nm$, the reflectivity is R = 0.15 and $d$ is the thickness of the layer in which the laser is absorbed. For $d = 1~\mu$m, the thickness of the film, the highest fluence of 12~mJ/cm$^2$ corresponds to a temperature rise of $\Delta T \approx 28$~K. This is just sufficient to raise the temperature of the film from 340~K to above $T_{MT}$, under the assumption that the heat is evenly distributed. This is in good agreement with the observed rocking curves where lower laser fluences only lead to a partial transformation.

\putbib
\end{bibunit}

\end{document}